\documentclass[aps,prl,reprint,superscriptaddress]{revtex4-1}

\usepackage[T1]{fontenc}
\usepackage[utf8]{inputenc}
\usepackage{graphicx}
\usepackage{amsmath}
\usepackage{amssymb}
\usepackage{xcolor}
\usepackage{pifont}

\newcommand{\beq}{\begin{equation}}
\newcommand{\eeq}{\end{equation}}

\begin{document}

\title{Universal Structural Relaxation in Supercooled Liquids}

\author{Florian Pabst}
\affiliation{Institute of Condensed Matter Physics, Technical University of Darmstadt, 64289 Darmstadt, Germany.}
\author{Jan Gabriel}
\affiliation{Institute of Condensed Matter Physics, Technical University of Darmstadt, 64289 Darmstadt, Germany.}
\author{Till Böhmer}
\affiliation{Institute of Condensed Matter Physics, Technical University of Darmstadt, 64289 Darmstadt, Germany.}
\author{Peter Weigl}
\affiliation{Institute of Condensed Matter Physics, Technical University of Darmstadt, 64289 Darmstadt, Germany.}
\author{Andreas Helbling}
\affiliation{Institute of Condensed Matter Physics, Technical University of Darmstadt, 64289 Darmstadt, Germany.}
\author{Timo Richter}
\affiliation{Institute of Condensed Matter Physics, Technical University of Darmstadt, 64289 Darmstadt, Germany.}
\author{Parvaneh Zourchang}
\affiliation{Institute of Condensed Matter Physics, Technical University of Darmstadt, 64289 Darmstadt, Germany.}
\author{Thomas Walther}
\affiliation{Institute of Applied Physics, Technical University of Darmstadt, 64289 Darmstadt, Germany.}
\author{Thomas Blochowicz}
\affiliation{Institute of Condensed Matter Physics, Technical University of Darmstadt, 64289 Darmstadt, Germany.}
\email{thomas.blochowicz@physik.tu-darmstadt.de}

\date{\today}

\begin{abstract}
  One of the hallmarks of molecular dynamics in deeply supercooled liquids is the non-exponential character of the relaxation functions. It has been a long standing issue if ``universal'' features govern the lineshape of glassy dynamics independent of any particular molecular structure or interaction.  In the paper, we elucidate this matter by giving a comprehensive comparison of the spectral shape of depolarized light scattering and dielectric data of deeply supercooled liquids. The light scattering spectra of very different systems, e.g. hydrogen bonding and van der Waals liquids but also ionic systems, almost perfectly superimpose and show a generic lineshape of the structural relaxation, approximately following a high frequency power law $\omega^{-1/2}$. However, the dielectric loss peak shows a more individual shape. In systems with low dipole moment generic behavior is also observed in the dielectric spectra, while in strongly dipolar liquids additional crosscorrelation contributions mask the generic structural relaxation.
\end{abstract}

\maketitle

On supercooling a liquid below its melting point the viscosity and corresponding relaxation times show a pronounced increase and eventually the liquid solidifies and forms a glass. Despite the many applications of supercooled and glassy materials, e.g. as household glasses or plastics,  a generally accepted theory of the liquid-glass transition is still lacking. Many attempts were made to elucidate quasi-universal features of the dynamics on approaching the glass transition and to disentangle generic features of glassy dynamics from properties due to particular interactions or other molecular peculiarities. One of the most important hallmarks of supercooled liquid dynamics, besides the tremendous ``super-Arrhenius'' slowing down occurring close to the glass transition temperature, is the fact that structural ($\alpha$-) relaxation deviates from a single exponential decay. This is observed, e.g., in the polarization response after switching off an external electric field across a supercooled polar material, and is documented by a tremendous body of broadband dielectric data on supercooled liquids, which have been collected over the years \cite{Boehmer1993a, Nielsen2009a}. The reported broad distributions of relaxation times are generally understood to reflect so called dynamic heterogeneities, i.e., local environments in the disordered material, where the dynamics differs from the average \cite{Sillescu1999a,Glotzer2000a,Richert2002a}. However, a detailed understanding of the lineshape of dynamic quantities like the dielectric, light scattering or NMR susceptibilities has been a major challenge \cite{Blochowicz2006c}. In an attempt to reveal universal relaxation features either scaling procedures have been suggested  \cite{Dixon1990a,Paluch1998a} or  quantitative parametrizations of the lineshape were pursued \cite{Blochowicz2006a,Brodin2007a,Gainaru2009a}, without leading to generally accepted success.

In most cases, the discussion of a temperature dependent line shape is based on broadband dielectric data, i.e., the dielectric permittivity $\hat\varepsilon(\omega)$, as this quantity is readily available over the full dynamic and temperature range of interest \cite{Lunkenheimer2000a}. When comparing different glass formers with respect to the main relaxation peak in the dielectric loss, the structural relaxation shows largely different behavior with the KWW stretching parameter of the corresponding correlation function $C(t)=\exp(-(t/\tau)^{\beta_\text{KWW}})$ typically varying in the range of $\beta_\text{KWW}=0.5\ldots 0.8$ in the majority of systems \cite{Paluch2016a}. Because of this variation, there have been many considerations connecting the $\alpha$-relaxation stretching to other features in the supercooled state like the fragility \cite{Boehmer1993a} or the timescale of secondary processes \cite{Ngai1998c}, while in simulation studies the $\alpha$-relaxation stretching is often thought to be related to individual aspects of the intermolecular potential, like its anharmonicity \cite{Bordat2004a, Bordat2007a}. In contrast to these observations there are theoretical model considerations that predict generic features of the structural relaxation, in particular a high-frequency $\chi''(\omega)\propto \omega^{-1/2}$ behavior of the dielectric loss, c.f.\ \cite{Dyre2005a, Dyre2005b} and references therein. And recently, it was suggested that the generic $\alpha$-relaxation features show up in certain \emph{simple} liquids, where dynamic susceptibilities of different experimental methods, like the dielectric loss, light scattering susceptibility or shear mechanical response may become identical \cite{Niss2018a}. Some experimental evidence for the latter comes from an analysis of shear mechanical data \cite{Bierwirth2017a}, i.e. translational dynamics. For reorientational motion as probed by dielectric spectroscopy the situation is far less clear \cite{Nielsen2009a}, and thus, despite some hints pointing towards a universal $\alpha$-relaxation shape, the overall picture remains confusing. But as we show in the following this matter can be much elucidated when results of different techniques are compared that all probe molecular reorientation. Especially, besides the common dielectric technique, we focus on depolarized dynamic light scattering and a \emph{local} dielectric probe, called triplet state solvation dynamics.

\begin{figure*}
  \includegraphics[width=.49\textwidth]{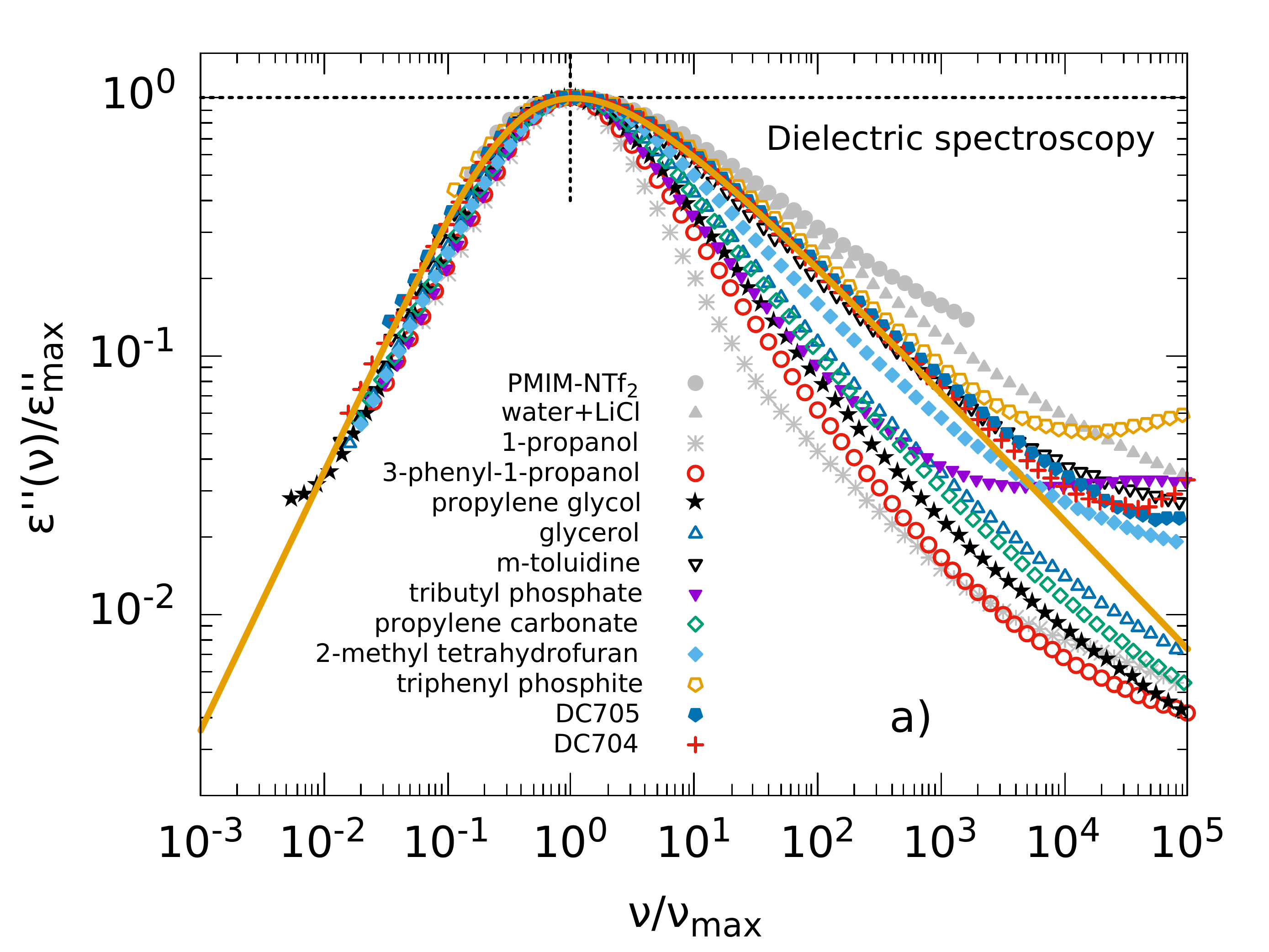}
  \includegraphics[width=.49\textwidth]{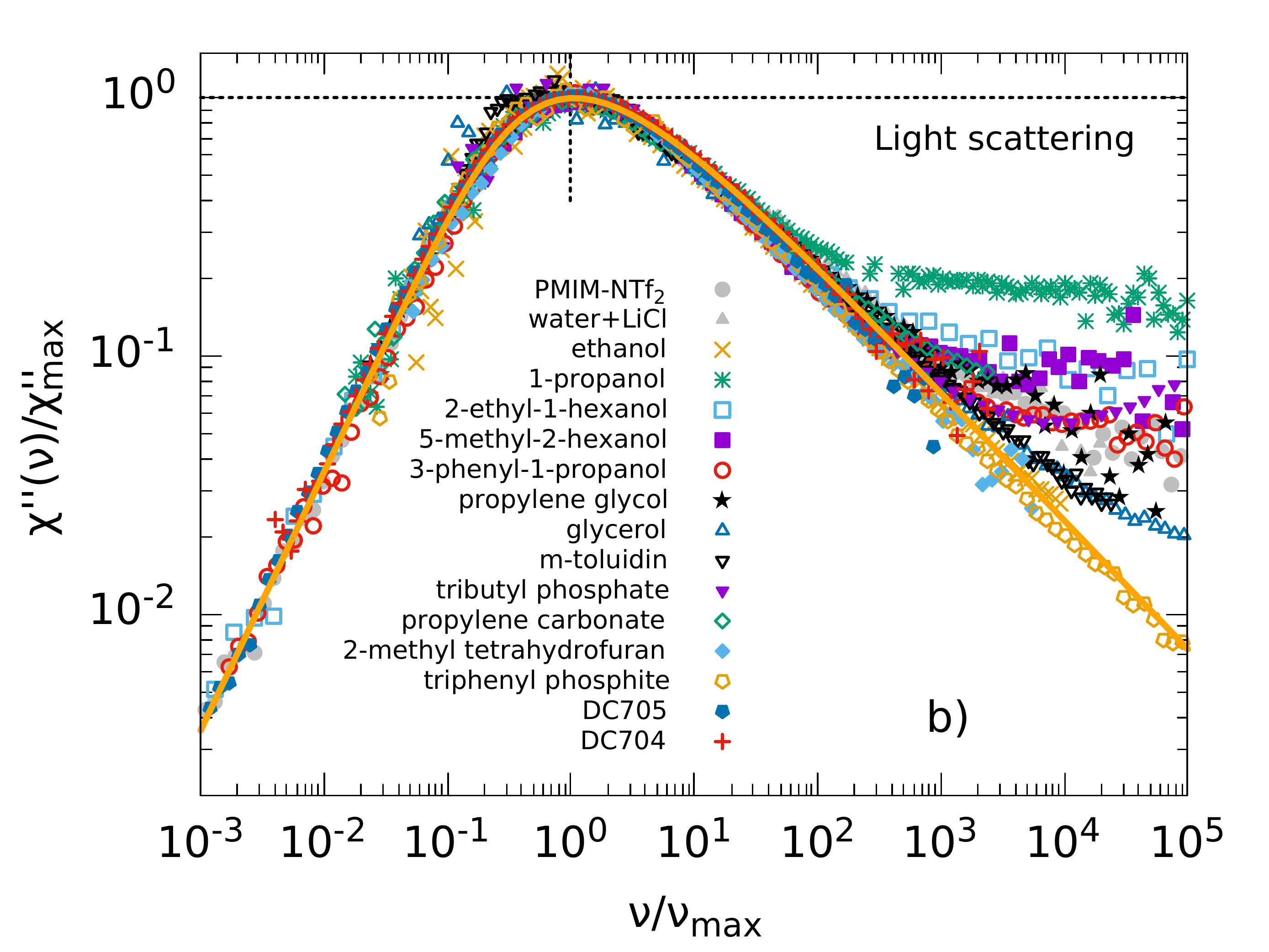}
  \caption{\label{fig:dls-master}  The lineshape of the $\alpha$-relaxation as probed by dielectric spectroscopy (left) and depolarized dynamic light scattering (right) in different substance classes. Solid line is a fit with high frequency power law $\omega^{-1/2}$. In the case of light scattering the data include besides monohydroxy alcohols, polyalcohols and van der Waals liquids also an ionic liquid, where the particular peak is known to represent molecular reorientation of cations \cite{Pabst2019a}, and also a LiCl-water mixture. In case of the dielectric data the ionic systems are shown in gray, as the latter contain additional ionic relaxation contributions. Moreover those monohydroxy alcohols apart from 1-propanol, for which there is consensus about a Debye like process in the literature were not included in the representation of the dielectric data for clarity. For references to data which are not from the present work see table \ref{tab:subst}.}
\end{figure*}
In Fig.~\ref{fig:dls-master} we show examples of a wide variety of different glass forming systems and compare the structural relaxation peak as obtained by dielectric spectroscopy (a) and depolarized light scattering (b). The chosen examples include substances with largely different polarities, and comprise van der Waals liquids as well as hydrogen bonding systems and also a few ionic systems, where it is known that light scattering reflects molecular reorientation, while dielectric loss also contains relaxation contributions of free charges \cite{Pabst2019a}.  It is well known, e.g., from Ref.~\cite{Paluch2016a}, that already for systems with just van der Waals interactions the high frequency slope of the dielectric  main relaxation peak largely varies in its power law exponent. This is reflected in the compilation of substances shown in Fig.~\ref{fig:dls-master}a), which do not superimpose beyond the low frequency part of the spectrum. Note that the ionic systems, for which the dielectric loss is hard to compare are shown in gray and those monohydroxy alcohols, for which there is consensus in the literature that an additional slow Debye process dominates the dielectric spectra, are not included in Fig.~\ref{fig:dls-master}a) for clarity, apart from 1-propanol that serves as an example for this substance class. But even though those substances were left out, no superposition is achieved.

\begin{table}
\footnotesize
\begin{tabular}{c|c|c|c|c|c}
  \textbf{substance} & $T_\text{BDS} /$K & $\Delta\varepsilon$ & $T_\text{DDLS} /$K & \textbf{ref.} & \\ \hline\hline
  PMIM-NTf2 (ion.\,liq.) &  200.0$^\ast$ & -- & 190.0$^\ast$ & \cite{Pabst2019a}&\hspace*{0em}\includegraphics[height=2ex]{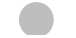}\hspace*{0em}\\ \hline
  17\% LiCl-water & 145.0 &--& 145.0 & -- &\hspace*{0em}\includegraphics[height=2ex]{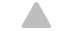}\hspace*{0em}\\ \hline\hline
  ethanol & \textcolor{gray}{97.0} & 117.5 & 97.0 & -- &\hspace*{0em}\includegraphics[height=2ex]{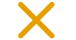}\hspace*{0em}\\ \hline
    1-propanol & 105.9 & 97.8 & 106.7$^\ast$ & \cite{Gabriel2017a}&\hspace*{0em}\includegraphics[height=2ex]{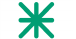}\hspace*{0em}\\ \hline
    2-ethyl-1-hexanol & \textcolor{gray}{170.3} & 25.8 & 152.9$^\ast$ & \cite{Gabriel2018a}&\hspace*{0em}\includegraphics[height=2ex]{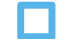}\hspace*{0em}\\\hline
    5-methyl-2-hexanol & \textcolor{gray}{170.3}& 26.3 & 160.0$^\ast$ & \cite{Gabriel2018a}&\hspace*{0em}\includegraphics[height=2ex]{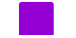}\hspace*{0em}\\\hline
    3-phenyl-1-propanol & 186.5 & 24.8 & 186.5 & \cite{Boehmer2019a}&\hspace*{0em}\includegraphics[height=2ex]{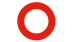}\hspace*{0em}\\ \hline\hline
    propylene glycol & 180.0 & 61.9 & 175.0 & -- &\hspace*{0em}\includegraphics[height=2ex]{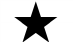}\hspace*{0em}\\ \hline
    glycerol & 200.0 & 68.4 & 190.0$^\ast$ &\cite{Gabriel2020a} &\hspace*{0em}\includegraphics[height=2ex]{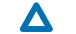}\hspace*{0em}\\ \hline
    m-toluidine & 190.7$^\ast$&6.9 & 185.0 & \cite{Benkhof1999a}&\hspace*{0em}\includegraphics[height=2ex]{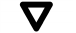}\hspace*{0em}\\ \hline
    tributyl phosphate & 147.3$^\ast$ &20.7 & 144.2$^\ast$ & \cite{Pabst2020a}&\hspace*{0em}\includegraphics[height=2ex]{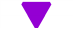}\hspace*{0em}\\ \hline
    propylene carbonate & 162.0 & 95.6 &162.0 & -- &\hspace*{0em}\includegraphics[height=2ex]{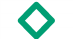}\hspace*{0em}\\ \hline
    2-methyl tetrahydrofuran & 98.0 & 14.8 & 91.9 & -- &\hspace*{0em}\includegraphics[height=2ex]{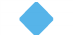}\hspace*{0em}\\ \hline
    triphenyl phosphite & 211.0 & 1.1 & 205.0 & -- &\hspace*{0em}\includegraphics[height=2ex]{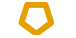}\hspace*{0em}\\ \hline
    DC705 & 235.2 & 0.2 & 245.2 & -- &\hspace*{0em}\includegraphics[height=2ex]{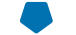}\hspace*{0em}\\ \hline
    DC704 & 220.0 & 0.2 & 225.0 & -- &\hspace*{0em}\includegraphics[height=2ex]{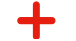}\hspace*{0em}\\ \hline
\end{tabular}
\caption{\label{tab:subst} Substances shown in Fig.~\ref{fig:dls-master} and temperatures used in the respective representation.$\,(^\ast)\,$indicates that data were taken from the given reference. Additionally the dielectric relaxation strength $\Delta\varepsilon$ at the respective temperature $T_\text{BDS}$ is given to provide a measure for the polarity.}
\end{table}
By contrast, for the corresponding light scattering spectra the case is entirely different. Here, besides all of the substances shown in a) also various monohydroxy alcohols are included, and Fig.~\ref{fig:dls-master}b) now demonstrates, that in the light scattering spectra indeed generic features of the $\alpha$-relaxation become obvious: despite the large differences in the dielectric counterparts, the light scattering spectra all show in good approximation the same spectral shape, which is well described by a high frequency power law $\propto\omega^{-1/2}$. In more detail, the fit shown as solid orange line in Fig.\ref{fig:dls-master} is based on a generalized gamma distribution of relaxation times as detailed in Ref.~\cite{Blochowicz2003a} with parameters $\alpha=2$ and $\beta=0.5$, which produces a peak width that lies in between that of the well known Kohlrausch stretched exponential and Cole-Davidson functions, both with a high frequency power law of $\omega^{-1/2}$. Of course deviations from this power law occur at higher frequencies, as further secondary relaxation features are present, the position and strength of which is obviously individual for each substance.

\begin{figure*}
  \includegraphics[width=\textwidth]{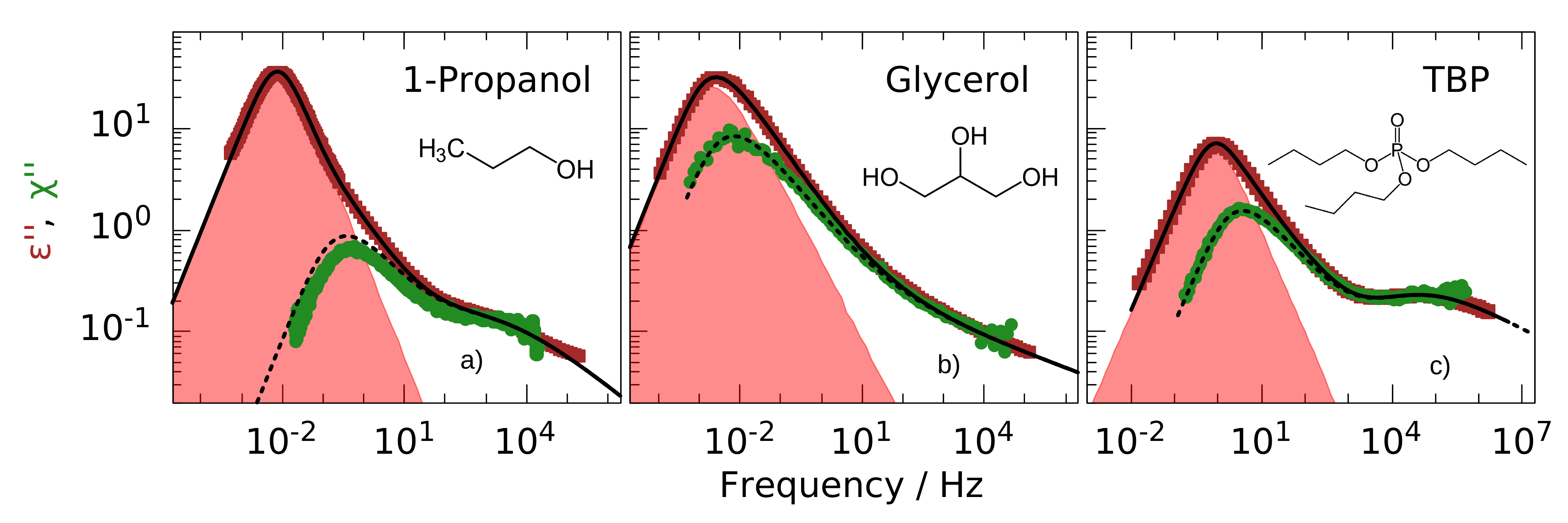}
  \caption{\label{fig:prop-gly-tbp} Examples of dielectric loss (red) and depolarized light scattering spectra (green) obtained at the same tamperature, for hydrogen bonding and non-hydrogen bonding polar systems. While in (a) the monohydroxy alcohol 1-propanol is shown (1 OH group), (b) represents the polyalcohol glycerol (3 OH) and in (c) tri-butyl-phosphate is shown (no OH). Data are taken from \cite{Gabriel2017a}, \cite{Gabriel2020a} and \cite{Pabst2020a}, respectively. The red shaded area represents the estimated contribution of dielectric crosscorrelations in each case. Solid lines represent the fit of the dielectric data set. Dashed lines are the contributions of $\alpha$- and $\beta$-relaxation to this fit.}
\end{figure*}
A starting point to understand the discrepancy between the spectra from dielectric spectroscopy and light scattering is the fact that different experimental methods largely differ in their sensitivity towards so-called crosscorrelations:
The macroscopic correlation or response functions obtained from experiments usually not only contain an autocorrelation part, say, of the molecular dipole moment $\boldsymbol\mu_i$ in case of broadband dielectric spectroscopy (BDS), but also crosscorrelation contributions between different molecules (dipole moments) $i\neq j$:
\begin{equation}
C_\text{BDS}(t) \propto   \sum_{i} \langle\boldsymbol\mu_i(0)\,\boldsymbol\mu_i(t)\rangle + \sum_{i\neq j} \langle\boldsymbol\mu_i(0)\,\boldsymbol\mu_j(t)\rangle,
\end{equation}
with $\langle\ldots\rangle$ denoting the statistical average. When searching for generic features of the structural relaxation, one would naturally refer to the autocorrelation part of the response, while crosscorrelations are expected to superimpose and possibly alter a generic lineshape. Crosscorrelations are usually considered for the static dielectric constant, where they are  quantified by the well-known Kirkwood correlation factor $g_K= 1 + \langle \boldsymbol\mu_i \sum_{i\neq j}\boldsymbol\mu_j\rangle/\mu^2 $, which proves to be $> 1$ for many dipolar substances \cite{Dejardin2020pre}.  Yet, their effect on the dynamics is only rarely considered. Recently, e.g., it was pointed out that the dipole moment plays a significant role in determining the spectral shape of the dielectric loss and a universal correlation between the dielectric relaxation strength $\Delta\varepsilon$ and the stretching parameter of the $\alpha$-process for 88 supercooled liquids was found\cite{Paluch2016a}, indicating that dipole-dipole correlations do play an important role for the dynamics in polar liquids \cite{Jedrzejowska2016a}.


While dielectric spectroscopy is particularly sensitive in that respect, other techniques, like, e.g., depolarized dynamic light scattering often provide more direct access to the self correlations and thus, to a possibly generic $\alpha$-relaxation. The latter technique probes molecular reorientation \cite{Berne1976a,Gabriel2018b}, not by fluctuations of the permanent dipole moment, as dielectric spectroscopy does, but by fluctuations of the molecular optical anisotropy. Therefore, dipole-dipole correlations play a smaller or even negligible role for light scattering, as was already demonstrated by experiments \cite{Boehmer2019a,Gabriel2018a}, and is now clearly indicated by the master curve in Fig.~\ref{fig:dls-master}b).

In order to understand the crosscorrelation effects in more detail, it is instructive to review a few examples, where the role of dipole-dipole correlations in the dielectric spectra was already investigated in some detail.
Good examples are found among the monohydroxy alcohols. As mentioned, the dielectric spectra in these systems are often dominated by crosscorrelations due to H-bonded supramolecular structures leading to the so-called \emph{Debye} process \cite{Boehmer2014a}. By contrast the light scattering spectra, if at all,  only show very weak signature of such a relaxation \cite{Gabriel2017a, Gabriel2018a}. This is exemplified for a dielectric and a light scattering spectrum of 1-propanol at $T = 106\,$K in Fig.~\ref{fig:prop-gly-tbp}(a): In the dielectric data (full squares) the Debye-peak (red shaded area) is superimposed with the structural ($\alpha$-) and secondary ($\beta$-) relaxations (dashed black line). The light scattering spectra (green circles), on the other hand,  are almost identical with the $\alpha$- and $\beta$-process of the dielectric fit, and thus prove to be insensitive to crosscorrelations \cite{Gabriel2017a}. Thus, the crosscorrelation contribution in the dielectric loss is easily identified.

It turns out that this procedure not only works for monohydroxy alcohols. In fact, in several polar substances, with or without hydrogen bonds, the dielectric spectrum at a given temperature can be represented as a sum of $\alpha$- and $\beta$-process from light scattering $\chi''^{DLS}_{\alpha\beta}(\nu)$ and a slow, Debye-like process representing the additional crosscorrelations. Fig.~\ref{fig:prop-gly-tbp} shows two more examples: first glycerol (b), a polyalcohol with three OH groups per molecule \cite{Gabriel2020a}  and  tributyl phosphate (c) a non-hydrogen bonding polar liquid \cite{Pabst2020a}. In the latter case the observation of an additional Debye-like process is particularly surprising, but in line with recent results by Dejardin et al., who expect from their calculations an additional slow relaxation process in dipolar liquids whenever $g_k>1$ \cite{Dejardin2019a}.


Although the analysis outlined in Fig.~\ref{fig:prop-gly-tbp} is not expected to work for all liquids, because depending on the exact motional mechanisms slight differences in strength and time constant may occur in the correlation functions of both methods \cite{Berne1976a,Gabriel2018b}, the results displayed in Fig.~\ref{fig:prop-gly-tbp} indicates two things: First, in molecular systems with large dipole moment the dielectric loss peak usually will not be dominated by the generic glassy dynamics contained in the self part of the correlation function, but will be superimposed by a strong crosscorrelation contribution. This, in turn, may also lead to a complex temperature dependence of the peak shape, as both contributions may have slightly different temperature dependencies. Thus, deviations from what is known as time-temperature superposition, i.e.\ the observation of a temperature independent line shape, which can be superimposed by simply scaling the time- or frequency axis, will naturally occur. Second, crosscorrelations will most likely be less important for light scattering, so that generic features of the lineshape will be easier visible in the light scattering than in the dielectric susceptibility. Of course the latter argument is limited to substances with not too large a structural anisotropy \cite{Battaglia1979a, Madden1980a}  and also for molecules with a pronounced chain-like character additional contributions are observed \cite{Ding2004a}. However, for cases similar to those represented in Fig.~\ref{fig:dls-master} both conclusions can be further tested starting from the ideas outlined in Fig.~\ref{fig:prop-gly-tbp}: If the above holds true then the light scattering and dielectric response function should give identical or at least very similar results in case of van der Waals liquids with low dipole moment and little structural anisotropy so that crosscorrelations in both experimental methods are negligible. 

\begin{figure}
  \includegraphics[width=.49\textwidth]{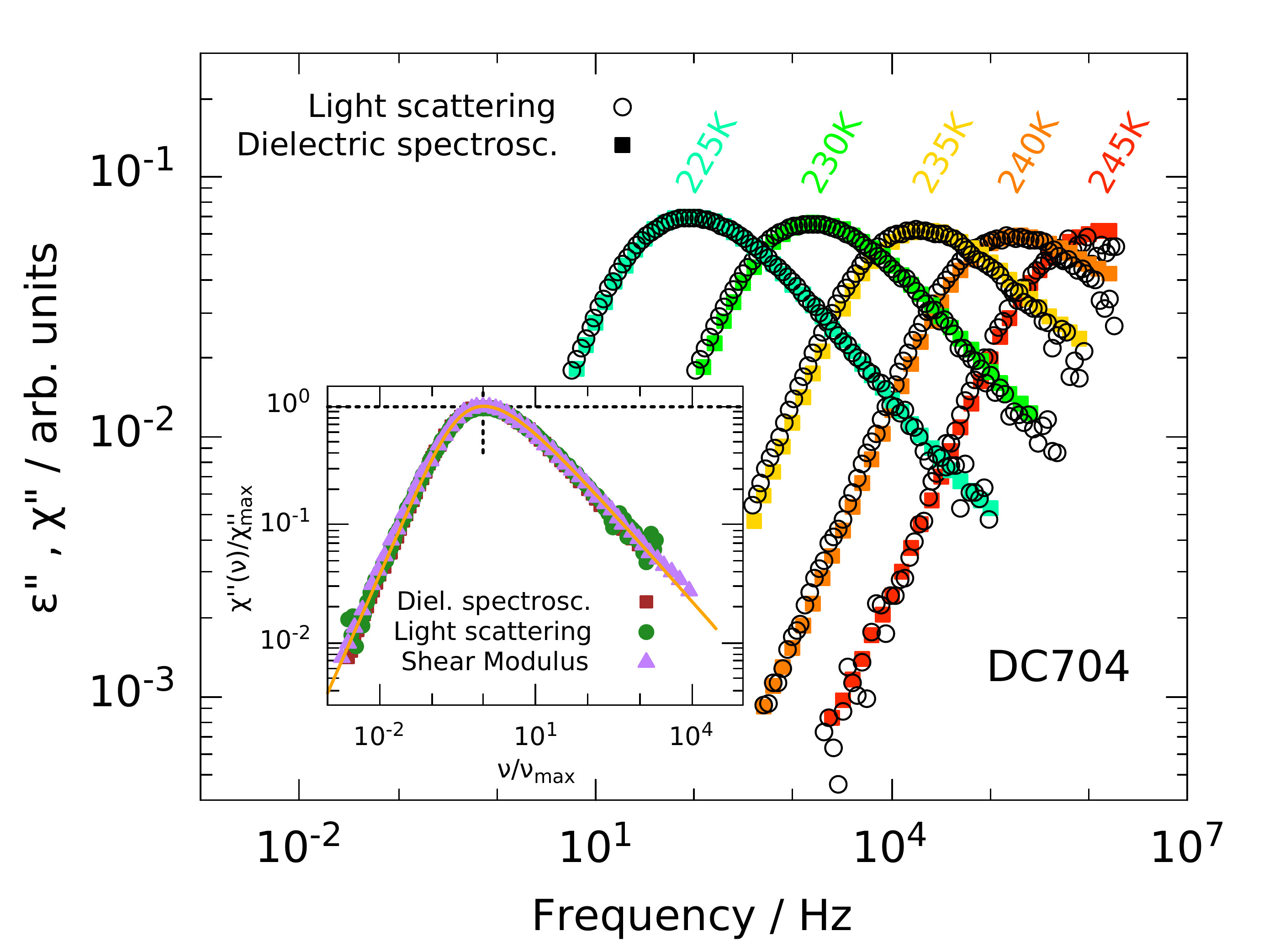}
  \caption{\label{fig:DC704} Detailed comparison of the light scattering and dielectric susceptibilities of the silicon oil DC704 with respective spectra obtained at the same temperatures. Only the absolute intensity of the light scattering spectra is scaled, i.e., both methods reveal the same dynamics due to lack of crosscorrelations. Inset: Dielectric and light scattering spectra together with shear modulus data from Ref.~\citenum{Jakobsen2005a} scaled to the peak maximum reveal the same spectral shape in all three methods. Solid line as in Figure \ref{fig:dls-master}.}
\end{figure}
For that purpose, we investigate the silicon oil DC 704 (tetramethyltetraphenyl trisiloxane), which is considered as one of the few examples of a \emph{simple liquid} in the literature \cite{Niss2018a}, i.e., a liquid that shows besides time-temperature superposition and density scaling, the same spectral shape in several observables and proportional time constants.     Measurements were done with dielectric spectroscopy and depolarized dynamic light scattering with particular care being taken to obtain data at identical temperatures in each technique. The result is shown in Fig.~\ref{fig:DC704}: the reorientational correlations probed by both methods are identical, including temperature dependence and lineshape of the susceptibility. Furthermore, when adding mechanical shear relaxation data and scaling the data onto the peak maximum (Fig.~\ref{fig:DC704} inset), all three techniques display the same lineshape, indicating that indeed the generic features of the $\alpha$-relaxation dominate the spectra. We note here that this holds true even though a modulus function (shear modulus) is compared with susceptibility functions (dielectric loss and light scattering susceptibility). In the present case of a liquid with low dipole moment this obviously does not affect the line shape.

Moreover, and not surprisingly, a fit reveals a high frequency power law $\chi''(\omega) \propto \omega^{-1/2}$ as expected from several theoretical models, independent of temperature, i.e., time-temperature superposition holds in good approximation. It should be noted, however, that in order to observe all of these features in the light scattering spectra it is required that the timescale of the $\alpha$-process is much longer than the microscopic dynamics in the THz regime, because otherwise further dynamic contributions overlap with the main relaxation peak. Thus, the generic features, like $\omega^{-1/2}$-behavior and time-temperature superposition are not observed in the dynamic regime of Tandem Fabry Perot interferometry (GHz -- THz) but rather in the one of photon correlation spectroscopy (mHz -- MHz), where the main relaxation is far enough separated from the microscopic relaxation features.

\begin{figure}
  \includegraphics[width=.49\textwidth]{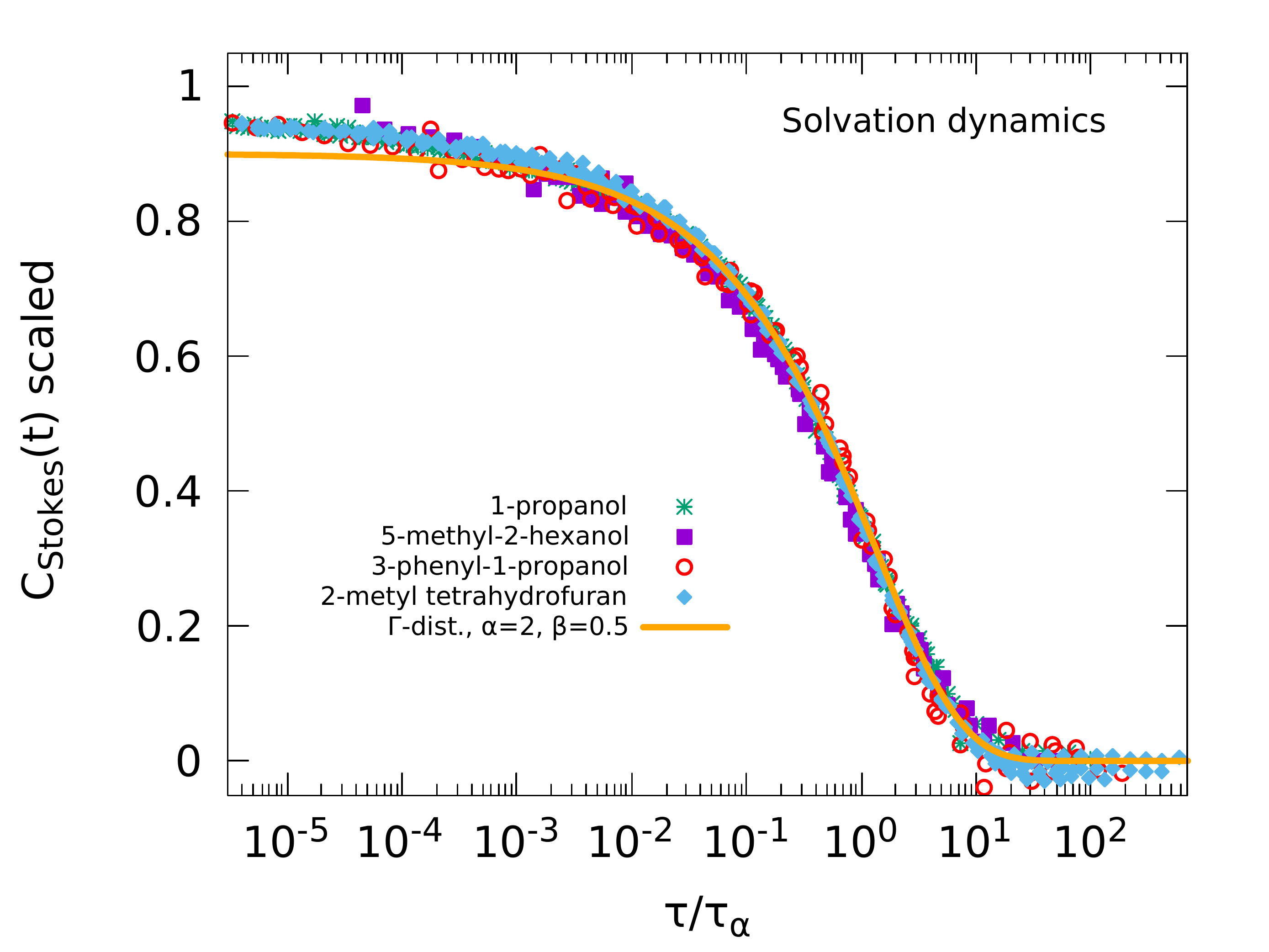}
  \caption{\label{fig:solvation-master} Rescaled relaxation functions of triplet state solvation dynamics with quinoxaline as phosphorescent dye molecule. In this case the method can be regarded as local dielectric spectroscopy. Propanol data from \cite{Weigl2019a}. Solid line: Relaxation function based on the same distribution of relaxation times as shown in Fig.~\ref{fig:dls-master}.}
\end{figure}
A further hint that indeed generic relaxation behavior is just covered by other dynamical features in BDS also comes from triplet state solvation dynamic experiments. This optical technique can under certain circumstances be regarded as local dielectric spectroscopy, collecting the dielectric response from a solvation shell of dipolar molecules around a phosphorescent dye \cite{Richert2000a}. Although many very local molecular properties are expected to play a role for the solvation response, it has recently been shown that this technique is rather insensitive towards the Debye process in monohydroxy alcohols and thus towards crosscorrelations \cite{Weigl2019a}. Fig.~\ref{fig:solvation-master} shows the $\alpha$-relaxation in a few systems, which obviously can be superimposed, showing a very similar lineshape despite the macroscopic dielectric counterparts being largely different due to different amounts of crosscorrelations. The solid line shown is the same lineshape as seen in Fig.~\ref{fig:dls-master} just Fourier transformed into the time domain.      


Thus, from the results presented above, a picture emerges that provides a new perspective on the universality issue: Indeed, there seems to be a generic structural relaxation that can be observed in deeply supercooled liquids and that shows a rather universal lineshape with a power law approximately $\propto \omega^{-1/2}$. However, direct observation may be hampered by several reasons: At higher temperatures the generic features are simply masked by other dynamic contributions and only become visible, when the liquid is deeply supercooled, so that generic behavior is possibly also a characteristic of the liquid state above $T_m$. Second, particularly in dielectric spectroscopy, generic behavior only becomes apparent in case of very low molecular dipole moments and in the absence of further particular interactions as in the presented example of DC704. With an increase in signal of the macroscopic dielectric loss, crosscorrelation contributions superimpose with the generic $\alpha$-relaxation of the self part of the correlation function, similar to the effects predicted by a recent theory \cite{Dejardin2019a}. These additional contributions lead to great variations in the apparent shape of the $\alpha$-peak and to natural deviations from time temperature superposition and also explain the correlations between relaxation strength and peak width reported in the literature \cite{Paluch2016a}. In depolarized light scattering and also in solvation dynamics, an extremely local dielectric probe, dipole correlations only play a minor role and the generic features appear more readily. Thus, a generic shape of the $\alpha$-relaxation seems to be present in a much larger number of systems than previously thought, which presents considerable challenge to most of present day glass transition theories that fall short of predicting such behavior \cite{Goetze2009a}, or are even in stark contrast to these observations \cite{Ngai1998b}, while other considerations \cite{Dyre2005a, Dyre2005b} seem to be supported by experimental evidence.
Therefore, dipole-dipole crosscorrelation contributions are not negligible for the dielectric response in the majority of dipolar systems, unlike often assumed, apart from cases of very low dipole moment.

\section{Method}
The dielectric data presented in this work were obtained with an Alpha-N high resolution analyzer by Novocontrol. Samples for the dielectric measurements were used as received. The light scattering setup used is described elsewhere in detail \cite{Gabriel2018b}. For the latter experiments all samples were filtered with 200\,nm syringe filters to reduce dust. The solvation experiment is described in detail elsewhere \cite{Weigl2018a}. The idea is that a dye molecule dissolved at low concentration is excited into a long lived triplet state. The time dependent Stokes shift of the emitted phosphorescent light of the dye reflects the time dependent polarization of the surrounding solvation shell and thus represents a local dielectric response of the material under study. For more details see \cite{Weigl2018a, Richert2000a}.

\section{Acknowledgements}
Financial support by the Deutsche Forschungsgemeinschaft under Grant
No. BL 1192/1 and BL 1192/3 is gratefully acknowledged. 

\bibliographystyle{achemso} 
\bibliography{glasses.bib}
   
\end{document}